\documentclass[superscriptaddress,aps,pra,twocolumn,nofootinbib,babel]{revtex4-1}
\usepackage{graphicx}
\usepackage{amsmath}
\usepackage{amssymb}
\usepackage{comment}
\usepackage{placeins}
\usepackage{rotating}
\usepackage[caption=false]{subfig}
\usepackage[english]{babel}
\usepackage{color}
\usepackage[normalem]{ulem}
\usepackage{algorithm}
\usepackage{algorithmic}
\usepackage{todonotes}
\usepackage{listings}
\usepackage[draft]{minted}

\makeatletter
\def\PYGdefault@reset{\let\PYGdefault@it=\relax \let\PYGdefault@bf=\relax%
    \let\PYGdefault@ul=\relax \let\PYGdefault@tc=\relax%
    \let\PYGdefault@bc=\relax \let\PYGdefault@ff=\relax}
\def\PYGdefault@tok#1{\csname PYGdefault@tok@#1\endcsname}
\def\PYGdefault@toks#1+{\ifx\relax#1\empty\else%
    \PYGdefault@tok{#1}\expandafter\PYGdefault@toks\fi}
\def\PYGdefault@do#1{\PYGdefault@bc{\PYGdefault@tc{\PYGdefault@ul{%
    \PYGdefault@it{\PYGdefault@bf{\PYGdefault@ff{#1}}}}}}}
\def\PYGdefault#1#2{\PYGdefault@reset\PYGdefault@toks#1+\relax+\PYGdefault@do{#2}}

\expandafter\def\csname PYGdefault@tok@gd\endcsname{\def\PYGdefault@tc##1{\textcolor[rgb]{0.63,0.00,0.00}{##1}}}
\expandafter\def\csname PYGdefault@tok@gu\endcsname{\let\PYGdefault@bf=\textbf\def\PYGdefault@tc##1{\textcolor[rgb]{0.50,0.00,0.50}{##1}}}
\expandafter\def\csname PYGdefault@tok@gt\endcsname{\def\PYGdefault@tc##1{\textcolor[rgb]{0.00,0.27,0.87}{##1}}}
\expandafter\def\csname PYGdefault@tok@gs\endcsname{\let\PYGdefault@bf=\textbf}
\expandafter\def\csname PYGdefault@tok@gr\endcsname{\def\PYGdefault@tc##1{\textcolor[rgb]{1.00,0.00,0.00}{##1}}}
\expandafter\def\csname PYGdefault@tok@cm\endcsname{\let\PYGdefault@it=\textit\def\PYGdefault@tc##1{\textcolor[rgb]{0.25,0.50,0.50}{##1}}}
\expandafter\def\csname PYGdefault@tok@vg\endcsname{\def\PYGdefault@tc##1{\textcolor[rgb]{0.10,0.09,0.49}{##1}}}
\expandafter\def\csname PYGdefault@tok@vi\endcsname{\def\PYGdefault@tc##1{\textcolor[rgb]{0.10,0.09,0.49}{##1}}}
\expandafter\def\csname PYGdefault@tok@vm\endcsname{\def\PYGdefault@tc##1{\textcolor[rgb]{0.10,0.09,0.49}{##1}}}
\expandafter\def\csname PYGdefault@tok@mh\endcsname{\def\PYGdefault@tc##1{\textcolor[rgb]{0.40,0.40,0.40}{##1}}}
\expandafter\def\csname PYGdefault@tok@cs\endcsname{\let\PYGdefault@it=\textit\def\PYGdefault@tc##1{\textcolor[rgb]{0.25,0.50,0.50}{##1}}}
\expandafter\def\csname PYGdefault@tok@ge\endcsname{\let\PYGdefault@it=\textit}
\expandafter\def\csname PYGdefault@tok@vc\endcsname{\def\PYGdefault@tc##1{\textcolor[rgb]{0.10,0.09,0.49}{##1}}}
\expandafter\def\csname PYGdefault@tok@il\endcsname{\def\PYGdefault@tc##1{\textcolor[rgb]{0.40,0.40,0.40}{##1}}}
\expandafter\def\csname PYGdefault@tok@go\endcsname{\def\PYGdefault@tc##1{\textcolor[rgb]{0.53,0.53,0.53}{##1}}}
\expandafter\def\csname PYGdefault@tok@cp\endcsname{\def\PYGdefault@tc##1{\textcolor[rgb]{0.74,0.48,0.00}{##1}}}
\expandafter\def\csname PYGdefault@tok@gi\endcsname{\def\PYGdefault@tc##1{\textcolor[rgb]{0.00,0.63,0.00}{##1}}}
\expandafter\def\csname PYGdefault@tok@gh\endcsname{\let\PYGdefault@bf=\textbf\def\PYGdefault@tc##1{\textcolor[rgb]{0.00,0.00,0.50}{##1}}}
\expandafter\def\csname PYGdefault@tok@ni\endcsname{\let\PYGdefault@bf=\textbf\def\PYGdefault@tc##1{\textcolor[rgb]{0.60,0.60,0.60}{##1}}}
\expandafter\def\csname PYGdefault@tok@nl\endcsname{\def\PYGdefault@tc##1{\textcolor[rgb]{0.63,0.63,0.00}{##1}}}
\expandafter\def\csname PYGdefault@tok@nn\endcsname{\let\PYGdefault@bf=\textbf\def\PYGdefault@tc##1{\textcolor[rgb]{0.00,0.00,1.00}{##1}}}
\expandafter\def\csname PYGdefault@tok@no\endcsname{\def\PYGdefault@tc##1{\textcolor[rgb]{0.53,0.00,0.00}{##1}}}
\expandafter\def\csname PYGdefault@tok@na\endcsname{\def\PYGdefault@tc##1{\textcolor[rgb]{0.49,0.56,0.16}{##1}}}
\expandafter\def\csname PYGdefault@tok@nb\endcsname{\def\PYGdefault@tc##1{\textcolor[rgb]{0.00,0.50,0.00}{##1}}}
\expandafter\def\csname PYGdefault@tok@nc\endcsname{\let\PYGdefault@bf=\textbf\def\PYGdefault@tc##1{\textcolor[rgb]{0.00,0.00,1.00}{##1}}}
\expandafter\def\csname PYGdefault@tok@nd\endcsname{\def\PYGdefault@tc##1{\textcolor[rgb]{0.67,0.13,1.00}{##1}}}
\expandafter\def\csname PYGdefault@tok@ne\endcsname{\let\PYGdefault@bf=\textbf\def\PYGdefault@tc##1{\textcolor[rgb]{0.82,0.25,0.23}{##1}}}
\expandafter\def\csname PYGdefault@tok@nf\endcsname{\def\PYGdefault@tc##1{\textcolor[rgb]{0.00,0.00,1.00}{##1}}}
\expandafter\def\csname PYGdefault@tok@si\endcsname{\let\PYGdefault@bf=\textbf\def\PYGdefault@tc##1{\textcolor[rgb]{0.73,0.40,0.53}{##1}}}
\expandafter\def\csname PYGdefault@tok@s2\endcsname{\def\PYGdefault@tc##1{\textcolor[rgb]{0.73,0.13,0.13}{##1}}}
\expandafter\def\csname PYGdefault@tok@nt\endcsname{\let\PYGdefault@bf=\textbf\def\PYGdefault@tc##1{\textcolor[rgb]{0.00,0.50,0.00}{##1}}}
\expandafter\def\csname PYGdefault@tok@nv\endcsname{\def\PYGdefault@tc##1{\textcolor[rgb]{0.10,0.09,0.49}{##1}}}
\expandafter\def\csname PYGdefault@tok@s1\endcsname{\def\PYGdefault@tc##1{\textcolor[rgb]{0.73,0.13,0.13}{##1}}}
\expandafter\def\csname PYGdefault@tok@dl\endcsname{\def\PYGdefault@tc##1{\textcolor[rgb]{0.73,0.13,0.13}{##1}}}
\expandafter\def\csname PYGdefault@tok@ch\endcsname{\let\PYGdefault@it=\textit\def\PYGdefault@tc##1{\textcolor[rgb]{0.25,0.50,0.50}{##1}}}
\expandafter\def\csname PYGdefault@tok@m\endcsname{\def\PYGdefault@tc##1{\textcolor[rgb]{0.40,0.40,0.40}{##1}}}
\expandafter\def\csname PYGdefault@tok@gp\endcsname{\let\PYGdefault@bf=\textbf\def\PYGdefault@tc##1{\textcolor[rgb]{0.00,0.00,0.50}{##1}}}
\expandafter\def\csname PYGdefault@tok@sh\endcsname{\def\PYGdefault@tc##1{\textcolor[rgb]{0.73,0.13,0.13}{##1}}}
\expandafter\def\csname PYGdefault@tok@ow\endcsname{\let\PYGdefault@bf=\textbf\def\PYGdefault@tc##1{\textcolor[rgb]{0.67,0.13,1.00}{##1}}}
\expandafter\def\csname PYGdefault@tok@sx\endcsname{\def\PYGdefault@tc##1{\textcolor[rgb]{0.00,0.50,0.00}{##1}}}
\expandafter\def\csname PYGdefault@tok@bp\endcsname{\def\PYGdefault@tc##1{\textcolor[rgb]{0.00,0.50,0.00}{##1}}}
\expandafter\def\csname PYGdefault@tok@c1\endcsname{\let\PYGdefault@it=\textit\def\PYGdefault@tc##1{\textcolor[rgb]{0.25,0.50,0.50}{##1}}}
\expandafter\def\csname PYGdefault@tok@fm\endcsname{\def\PYGdefault@tc##1{\textcolor[rgb]{0.00,0.00,1.00}{##1}}}
\expandafter\def\csname PYGdefault@tok@o\endcsname{\def\PYGdefault@tc##1{\textcolor[rgb]{0.40,0.40,0.40}{##1}}}
\expandafter\def\csname PYGdefault@tok@kc\endcsname{\let\PYGdefault@bf=\textbf\def\PYGdefault@tc##1{\textcolor[rgb]{0.00,0.50,0.00}{##1}}}
\expandafter\def\csname PYGdefault@tok@c\endcsname{\let\PYGdefault@it=\textit\def\PYGdefault@tc##1{\textcolor[rgb]{0.25,0.50,0.50}{##1}}}
\expandafter\def\csname PYGdefault@tok@mf\endcsname{\def\PYGdefault@tc##1{\textcolor[rgb]{0.40,0.40,0.40}{##1}}}
\expandafter\def\csname PYGdefault@tok@err\endcsname{\def\PYGdefault@bc##1{\setlength{\fboxsep}{0pt}\fcolorbox[rgb]{1.00,0.00,0.00}{1,1,1}{\strut ##1}}}
\expandafter\def\csname PYGdefault@tok@mb\endcsname{\def\PYGdefault@tc##1{\textcolor[rgb]{0.40,0.40,0.40}{##1}}}
\expandafter\def\csname PYGdefault@tok@ss\endcsname{\def\PYGdefault@tc##1{\textcolor[rgb]{0.10,0.09,0.49}{##1}}}
\expandafter\def\csname PYGdefault@tok@sr\endcsname{\def\PYGdefault@tc##1{\textcolor[rgb]{0.73,0.40,0.53}{##1}}}
\expandafter\def\csname PYGdefault@tok@mo\endcsname{\def\PYGdefault@tc##1{\textcolor[rgb]{0.40,0.40,0.40}{##1}}}
\expandafter\def\csname PYGdefault@tok@kd\endcsname{\let\PYGdefault@bf=\textbf\def\PYGdefault@tc##1{\textcolor[rgb]{0.00,0.50,0.00}{##1}}}
\expandafter\def\csname PYGdefault@tok@mi\endcsname{\def\PYGdefault@tc##1{\textcolor[rgb]{0.40,0.40,0.40}{##1}}}
\expandafter\def\csname PYGdefault@tok@kn\endcsname{\let\PYGdefault@bf=\textbf\def\PYGdefault@tc##1{\textcolor[rgb]{0.00,0.50,0.00}{##1}}}
\expandafter\def\csname PYGdefault@tok@cpf\endcsname{\let\PYGdefault@it=\textit\def\PYGdefault@tc##1{\textcolor[rgb]{0.25,0.50,0.50}{##1}}}
\expandafter\def\csname PYGdefault@tok@kr\endcsname{\let\PYGdefault@bf=\textbf\def\PYGdefault@tc##1{\textcolor[rgb]{0.00,0.50,0.00}{##1}}}
\expandafter\def\csname PYGdefault@tok@s\endcsname{\def\PYGdefault@tc##1{\textcolor[rgb]{0.73,0.13,0.13}{##1}}}
\expandafter\def\csname PYGdefault@tok@kp\endcsname{\def\PYGdefault@tc##1{\textcolor[rgb]{0.00,0.50,0.00}{##1}}}
\expandafter\def\csname PYGdefault@tok@w\endcsname{\def\PYGdefault@tc##1{\textcolor[rgb]{0.73,0.73,0.73}{##1}}}
\expandafter\def\csname PYGdefault@tok@kt\endcsname{\def\PYGdefault@tc##1{\textcolor[rgb]{0.69,0.00,0.25}{##1}}}
\expandafter\def\csname PYGdefault@tok@sc\endcsname{\def\PYGdefault@tc##1{\textcolor[rgb]{0.73,0.13,0.13}{##1}}}
\expandafter\def\csname PYGdefault@tok@sb\endcsname{\def\PYGdefault@tc##1{\textcolor[rgb]{0.73,0.13,0.13}{##1}}}
\expandafter\def\csname PYGdefault@tok@sa\endcsname{\def\PYGdefault@tc##1{\textcolor[rgb]{0.73,0.13,0.13}{##1}}}
\expandafter\def\csname PYGdefault@tok@k\endcsname{\let\PYGdefault@bf=\textbf\def\PYGdefault@tc##1{\textcolor[rgb]{0.00,0.50,0.00}{##1}}}
\expandafter\def\csname PYGdefault@tok@se\endcsname{\let\PYGdefault@bf=\textbf\def\PYGdefault@tc##1{\textcolor[rgb]{0.73,0.40,0.13}{##1}}}
\expandafter\def\csname PYGdefault@tok@sd\endcsname{\let\PYGdefault@it=\textit\def\PYGdefault@tc##1{\textcolor[rgb]{0.73,0.13,0.13}{##1}}}

\def\PYGdefaultZus{\char`\_}
\def\PYGdefaultZob{\char`\{}
\def\PYGdefaultZcb{\char`\}}

\def\PYGdefaultZhy{\char`\-}
\def\PYGdefaultZsq{\char`\'}
\def\PYGdefaultZdq{\char`\"}

\makeatother

\newcommand{\eq}[1]{Eq.~\hyperref[eq:#1]{(\ref*{eq:#1})}}
\renewcommand{\sec}[1]{\hyperref[sec:#1]{Section~\ref*{sec:#1}}}
\newcommand{\app}[1]{\hyperref[app:#1]{Appendix~\ref*{app:#1}}}
\newcommand{\tab}[1]{\hyperref[tab:#1]{Table~\ref*{tab:#1}}}
\newcommand{\fig}[1]{\hyperref[fig:#1]{Figure~\ref*{fig:#1}}}
\newcommand{\figa}[2]{\hyperref[fig:#1]{Figure~\ref*{fig:#1}#2}}
\newcommand{\figx}[2]{\hyperref[fig:#1]{Figure~\ref*{fig:#1}(#2)}}
\newcommand{\thm}[1]{\hyperref[thm:#1]{Theorem~\ref*{thm:#1}}}
\newcommand{\lem}[1]{\hyperref[lem:#1]{Lemma~\ref*{lem:#1}}}
\newcommand{\cor}[1]{\hyperref[cor:#1]{Corollary~\ref*{cor:#1}}}
\newcommand{\defn}[1]{\hyperref[def:#1]{Definition~\ref*{def:#1}}}
\newcommand{\alg}[1]{\hyperref[alg:#1]{Algorithm~\ref*{alg:#1}}}

\newcommand{\ignore}[1]{}

\newcommand{\be}{\begin{equation}}
\newcommand{\ee}{\end{equation}}
\newcommand{\ba}{\begin{eqnarray}}
\newcommand{\ea}{\end{eqnarray}}
\usepackage{times}

\begin{document}

\title{Validating Quantum-Classical Programming Models with Tensor Network Simulations}

\author{Alexander McCaskey}
\email[Corresponding author: ]{mccaskeyaj@ornl.gov}
\affiliation{Quantum Computing Institute, Oak Ridge National Laboratory, Oak Ridge, Tennessee 37831, United States}
\affiliation{Computer Science and Mathematics Division, Oak Ridge National Laboratory, Oak Ridge, Tennessee 37831, United States}

\author{Eugene Dumitrescu}
\affiliation{Quantum Computing Institute, Oak Ridge National Laboratory, Oak Ridge, Tennessee 37831, United States}
\affiliation{Computational Sciences and Engineering Division, Oak Ridge National Laboratory, Oak Ridge, Tennessee 37831, United States}

\author{Mengsu Chen}
\affiliation{Department of Physics, Virginia Tech, Blacksburg, Virginia 24060, United States}

\author{Dmitry Lyakh}
\affiliation{Quantum Computing Institute, Oak Ridge National Laboratory, Oak Ridge, Tennessee 37831, United States}
\affiliation{National Center for Computational Sciences, Oak Ridge National Laboratory, Oak Ridge, Tennessee 37831, United States}

\author{Travis S.~Humble}
\affiliation{Quantum Computing Institute, Oak Ridge National Laboratory, Oak Ridge, Tennessee 37831, United States}
\affiliation{Computational Sciences and Engineering Division, Oak Ridge National Laboratory, Oak Ridge, Tennessee 37831, United States}
\affiliation{Bredesen Center for Interdisciplinary Research, University of Tennessee, Knoxville, Tennessee 37996, United States}

\thanks{This manuscript has been authored by UT-Battelle, LLC, under Contract No.~DE-AC0500OR22725 with the U.S.~Department of Energy. The United States Government retains and the publisher, by accepting the article for publication, acknowledges that the United States Government retains a non-exclusive, paid-up, irrevocable, world-wide license to publish or reproduce the published form of this manuscript, or allow others to do so, for the United States Government purposes. The Department of Energy will provide public access to these results of federally sponsored research in accordance with the DOE Public Access Plan.}

\date{\today}

\begin{abstract}
The exploration of hybrid quantum-classical algorithms and programming models on noisy near-term quantum hardware has begun. As hybrid programs scale towards classical intractability, validation and benchmarking are critical to understanding the utility of the hybrid computational model. In this paper, we demonstrate a newly developed quantum circuit simulator based on tensor network theory that enables intermediate-scale verification and validation of hybrid quantum-classical computing frameworks and programming models. We present our tensor-network quantum virtual machine (TNQVM) simulator which stores a multi-qubit wavefunction in a compressed (factorized) form as a matrix product state, thus enabling single-node simulations of larger qubit registers, as compared to brute-force state-vector simulators. Our simulator is designed to be extensible in both the tensor network form and the classical hardware used to run the simulation (multicore, GPU, distributed). The extensibility of the TNQVM simulator with respect to the simulation hardware type is achieved via a pluggable interface for different numerical backends (e.g., ITensor and ExaTENSOR numerical libraries). We demonstrate the utility of our TNQVM quantum circuit simulator through the verification of randomized quantum circuits and the variational quantum eigensolver algorithm, both expressed within the eXtreme-scale ACCelerator (XACC) programming model.
\end{abstract}

\maketitle

\section{Introduction}\label{sec:intro}
Quantum computing is a computation paradigm that relies on the principles of quantum mechanics in order to process information. Recent advances in both algorithmic research, which has found remarkable speed-ups for a growing number of applications \cite{Childs2010,Montanaro2016, Biamonte2017}, and hardware development \cite{Linke2017,Friis2018} continue to progress the field of quantum information processing. The near-term state of quantum computing is defined by the noisy intermediate-scale quantum (NISQ) paradigm which involves small-scale noisy quantum processors \cite{preskillNISQ2018} being used in a hybrid quantum-classical framework. In this context, recent experimental demonstrations \cite{Peruzzo2014, O'Malley2016, Kandala2017, otterbach2017, deuteron} of hybrid computations have reinforced the need for robust programming models and classical validation frameworks.
\par
The successful integration of quantum processors into conventional computational workloads is a complex task which depends on the programming and execution models that define how quantum resources interact with conventional computing systems \cite{Humble2016HPEC,Britt2017}. Many different models have been proposed for programming quantum computers and a number of software development efforts have begun focusing on high-level hybrid programming mechanisms capable of integrating both conventional and quantum computing processors together \cite{Green2013,Javadiabhari2014,Wecker2014,Humble2014,smith2016practical,liu2017q,Svore2018,pakin_2018}. For example, recent efforts have focused on Python-based programming frameworks enabling the high-level expression of quantum programs in a classical context, which may target numerical simulators or a variety of physical quantum processing units (QPUs) \cite{1608.03355, projectq, qiskit}. The eXtreme-scale ACCelerator programming model (XACC) is a recently-developed quantum-classical programming, compilation, and execution framework that enables programming across multiple languages and targets multiple virtual and physical QPUs \cite{xaccarxiv}.
\par
In all cases, the verification of quantum program correctness is a challenging and complex task due to the intrinsically noisy nature of near-term QPUs, and this is additionally complicated by remote hosting. As a remedy, numerical simulation techniques can greatly expedite the analysis of quantum-classical programming efforts by providing direct insight into the prepared quantum states, as well as serving to test a variety of quantum computing hardware models. Modeling and simulation is essential for designing effective program execution mechanisms because it provides a controlled environment for understanding how complex computational systems interact, subsequently generating feedback based on the state machine statistics. For example, the performance of existing QPUs is limited by the hardware connectivity \cite{Linke2017} and numerical simulations can draw on a broad range of parameterized models to test new processor layouts and architectures.
\par
In practice, exact brute-force simulations of quantum computing are notoriously inefficient in memory complexity due to the exponential growth in resources with respect to system size. These brute-force methods explicitly solve the Schrodinger equation, or a mixed-state master equation, using a full representation of the quantum state in its underlying (exponentially large) Hilbert space. Limits on available memory place upper bounds on the size of the vectors or density operators that can physically be stored, severely restricting the size the simulated quantum circuit. Even with the availability of current large-scale HPC systems, including the state-of-the-art supercomputing systems, recent records for quantum circuit simulations are limited to less than 50 qubits \cite{nersc45, Pednault2017}. The performance of the brute-force quantum circuit simulators on current supercomputing architectures is also limited by the inherently low arithmetic intensity (Flop/Byte ratio) of the underlying vector operations (sparse matrix-vector multiplications) required for simulating a discrete sequence of one- and two-qubit gates.
\par
The inherent inefficiency of the brute-force state-vector quantum circuit simulators has motivated a search for approximate numerical simulation techniques increasing the upper bound on the number of simulated qubits. As we are interested in general-purpose (universal) quantum circuit simulators, we will omit efficient specialized simulation algorithms that target certain subclasses of quantum circuits, for example, quantum circuits composed of only Clifford operations \cite{Aaronson2004}. As a general solution, we advocate for the use of the tensor network (TN) theory as a tool for constructing factorized approximations to the exact multi-qubit wave-function tensor. The two main advantages offered by the tensor-network based wave-function factorization are (1) the memory (space) and time complexity of the quantum circuit simulation reflect the level of entanglement in the quantum system, (2) the numerical action of quantum gates on the factorized wave-function representation results in numerical operations (tensor contractions) which become arithmetically intensive for entangled systems, thus potentially delivering close to the peak utilization of modern HPC platforms.
\par
\section{Quantum Circuit Simulation with Tensor Networks}
Tensor network theory \cite{Orus2014,Biamonte2017} provides a versatile and modular approach for the dimensionality reduction of operators acting in high-dimensional linear spaces. For the following discussion, a tensor is a generalization of a vector and is defined in a linear space constructed as the tensor product of two or more primitive vector spaces. Consequently, the components of a tensor are enumerated by a tuple of indices, instead of by a single index as is the case for vectors. From the numerical perspective, a tensor can be viewed as a multi-dimensional array of objects, which may be real or complex numbers. In this work, following the physics nomenclature, we shall refer to the number of indices in a tensor \(T_{i_1 ... i_n}\) as its rank (in this case the tensor rank is $n$). Each index represents a distinct vector space contributing to the composite space by the tensor product. The extent of the range of each index gives the dimension of the vector space. In their essence, tensor networks aim at decomposing higher-rank tensors into a contraction over lower-rank tensors such that the factorized product accurately reconstructs properties of the original tensor (i.e. a variant of lossy compression in linear spaces). Any tensor can be approximated by a suitably chosen tensor network with arbitrary precision, however the size of the tensor factors may grow exponentially in worst case examples. Tensor factorizations, which we also refer to as decompositions, are not unique in general and the problem of finding the optimal tensor decomposition is a difficult non-convex optimization problem \cite{Kolda2015}.

In practice, a tensor network factorization is typically specified by a graph in which the nodes are the tensor factors and the edges represent physical or auxiliary vector spaces which are associated with the indices of the corresponding tensor factors. A closed edge, that is, an edge connecting two nodes, represents a contracted index shared by two tensor factors over which a summation is to be performed. In a standalone tensor network, contracted indices are associated with auxiliary vector spaces. An open edge, that is, an edge connected to only one node, represents an uncontracted index of that tensor factor. Uncontracted indices are typically associated with physical vector spaces. Different tensor network architectures differ by the topology of their representative graphs. Furthermore, one can define even more general tensor network architectures by replacing graphs with hypergraphs, in which case an edge may connect three or more tensors. In the subsequent discussion, however, we will mostly deal with conventional graph topologies.
\begin{figure}[ht!]
\centering
\includegraphics[width=\columnwidth]{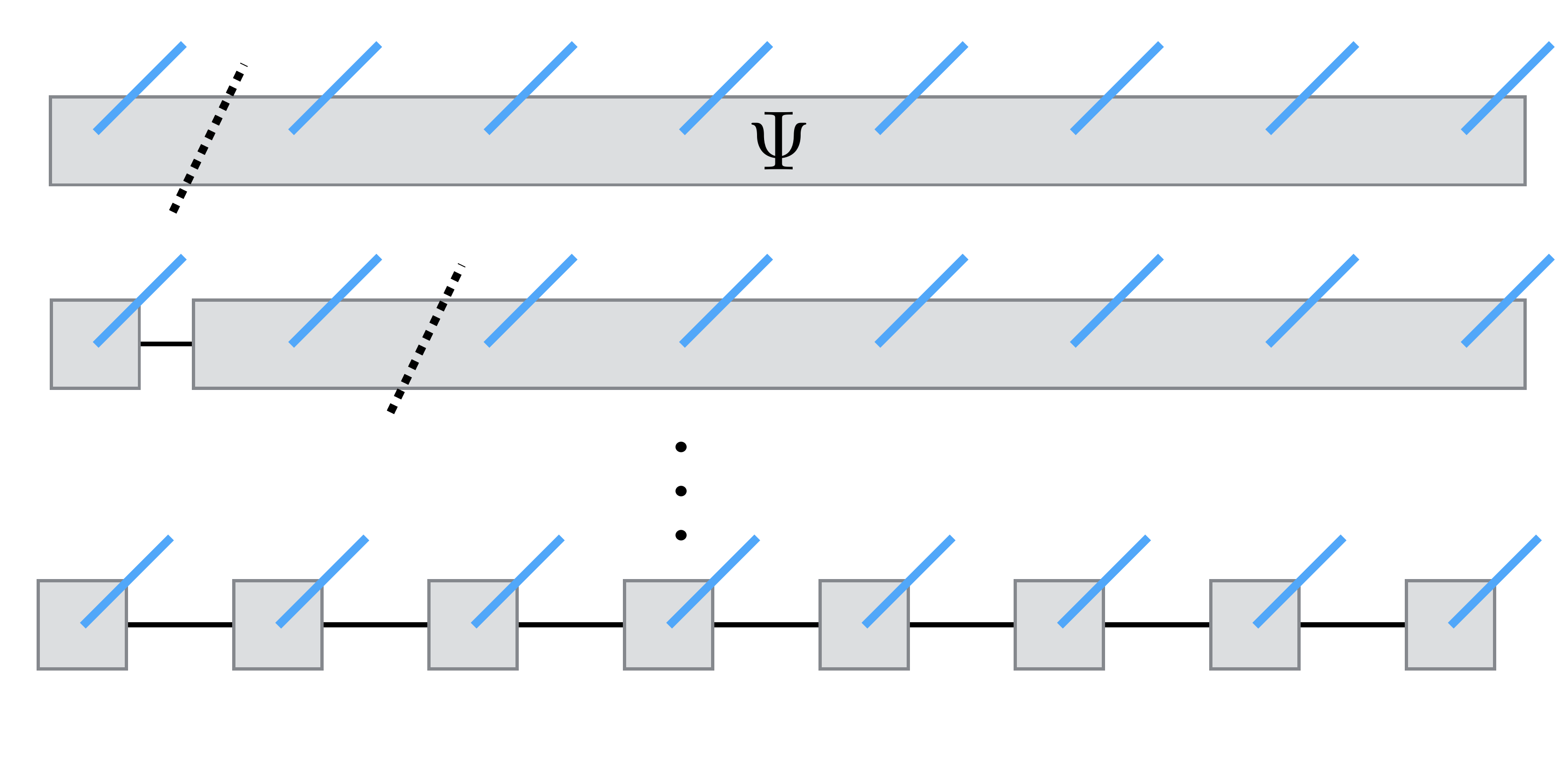}
\caption{Figure depicting the tensor decomposition of a wavefunction into the MPS form}
\label{fig:tn}
\end{figure}

\par
A quantum many-body wave-function, including a multi-qubit wave-function, is essentially a high-rank tensor (its rank is equal to the number of simulated quantum particles or quasi-particles) \cite{Orus2014}. A number of different tensor network architectures have been suggested for the purpose of factorizing quantum many-body wave-functions, including the matrix-product state (MPS) \cite{White:1992ie, Schollwock2011}, the projected entangled pair-state (PEPS) \cite{Schuch2007,Verstraete2008}, the tree tensor network state (TTNS) \cite{Murg2010, Nakatani2013, Dumitrescu2017}, the multiscale entanglement renormalization ansatz (MERA) \cite{vidal2006, evenbly2009algorithms}, as well as somewhat related non-conventional schemes like the complete-graph tensor network (CGTN) \cite{cgtn}. All of the above tensor network \textit{ansaetze} differ in the factorization topology, that is, in how the tensor factors are contracted with each other to form the final quantum many-body wave-function. In a good tensor network factorization, topology is induced by the entanglement structure of a particular quantum many-body system. Many physical systems are described by many-body Hamiltonians with only local interactions -- in many cases, nearest neighbor only -- with correlation functions decaying exponentially for non-critical states. In such cases, the locality structure of the many-body Hamiltonian induces the necessary topology required to properly capture the quantum correlations present in the system of interest. The factorization topology also strongly affects the computational cost associated with the numerical evaluation/optimization of a specific tensor network architecture. Another important characteristic of a tensor network is its so-called maximal bond dimension, that is, the maximal dimension of the auxiliary linear spaces (auxiliary linear spaces are those contracted over). Provided that the maximal bond dimension is bounded, many tensor network factorizations can be evaluated with a polynomial computational cost in the bond dimension. In practice, the entanglement structure of the underlying quantum many-body system determines the maximal bond dimension needed for a given error tolerance and a given tensor network topology. A poorly chosen tensor network topology will necessarily lead to rapidly increasing (exponentially at worst) bond dimensions in order to keep the factorization error within the error threshold.

The entanglement structure in a multi-qubit wave-function is determined by the quantum circuit and may be unknown in general. Consequently, there is no well-defined choice of a tensor network architecture (topology) that could work equally well for all quantum circuits, unless it is some kind of an adaptive topology. In practice, the choice of a tensor network architecture for representing a multi-qubit wave-function is often dictated by numerical convenience and ease of implementation. For example, one of the simplest tensor network architectures, the MPS ansatz, was used to simulate Shor's algorithm for integer factorization \cite{Dang2017}. Although the inherently one-dimensional chain topology of the MPS ansatz often results in severely growing bond dimensions, and this can be remedied by a more judicious tensor network form\cite{Dumitrescu2017}, its computational convenience and well understood theory makes the MPS factorization an appealing first candidate for our quantum virtual machine (quantum circuit simulator). In future, we plan on adding more advanced tensor network architectures.

\begin{figure}[ht!]
\centering
\includegraphics[width=\columnwidth]{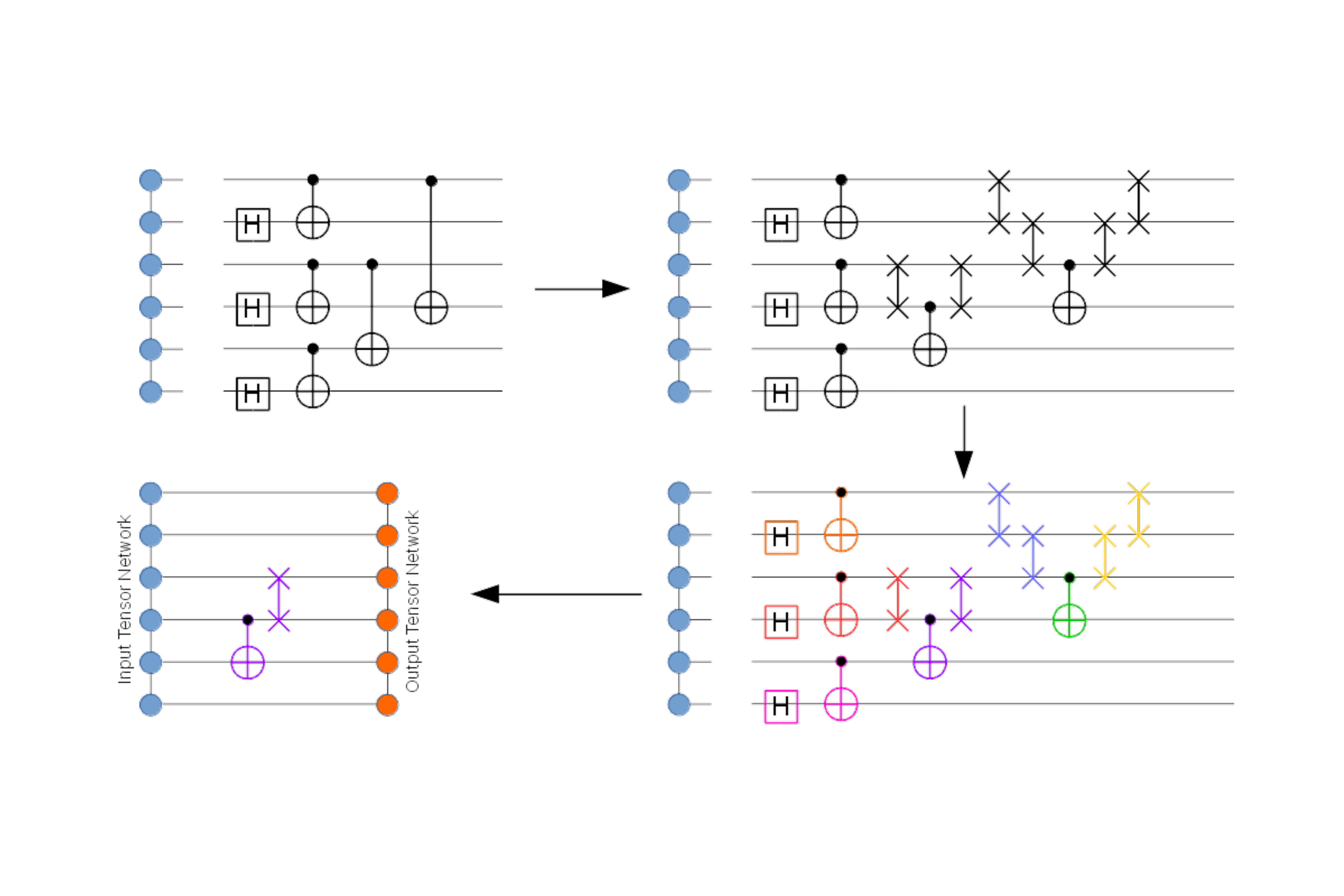}
\caption{Graphical illustration of the general quantum circuit simulation algorithm with the qubit wave-function factorized as a tensor network.}
\label{fig:tnalg}
\end{figure}

In order to simulate a general quantum circuit over an $N$-qubit register with the tensor network machinery the following steps will be necessary (see Figure \ref{fig:tnalg}):
\begin{enumerate}
    \item Specify the chosen tensor network graph that factorizes the rank-$N$ wave-function tensor into a contracted product of lower-order tensors (factors).
    \item Transform the quantum circuit into an equivalent quantum circuit augmented with SWAP gates in order to maximize the number of accelerated gate applications (see below). This is an optional step.
    \item Group quantum gates into ordered aggregates (super-gates) which will act as a whole on the qubit wave-function. In the simplest case, all quantum gates will be applied one-by-one in order of appearance, with no aggregation. This is an optional step.
    \item Sequentially apply aggregated super-gates (or individual gates when no aggregation occurred) to the wave-function tensor network, thus evolving it towards the output state.
\end{enumerate}

In the above general algorithm, the application of a super-gate (or just an individual gate) on a multi-qubit wave-function tensor consists of the following steps:
\begin{enumerate}
    \item Append the individual gates constituting the given super-gate to the input wave-function tensor network \(TN_{inp}\), thus obtaining a larger tensor network \(TN_{mid}\).
    \item If there are 2- or higher-body gates present, check whether they are applied to the qubit pairs or triples, etc. that allow accelerated gate application (for example, in MPS factorization, these would be the adjacent qubit pairs, triples, and so on). If yes, evaluate their action in an accelerated fashion (see below). Otherwise, resort to the general algorithm in the next steps.
    \item Instantiate a new tensor network \(TN_{out}\) by cloning \(TN_{inp}\).
    \item Close \(TN_{mid}\) with \(TN_{out}\), thus obtaining a closed tensor network \(TN_{opt}\).
    \item Optimize the tensors of \(TN_{out}\) to maximize \(TN_{opt}\).
    \item If \(TN_{opt}\) value is not acceptable, increase dimensions of the auxiliary spaces in \(TN_{out}\) and repeat Step 5.
\end{enumerate}

\begin{figure}[ht!]
\centering
\includegraphics[width=\columnwidth]{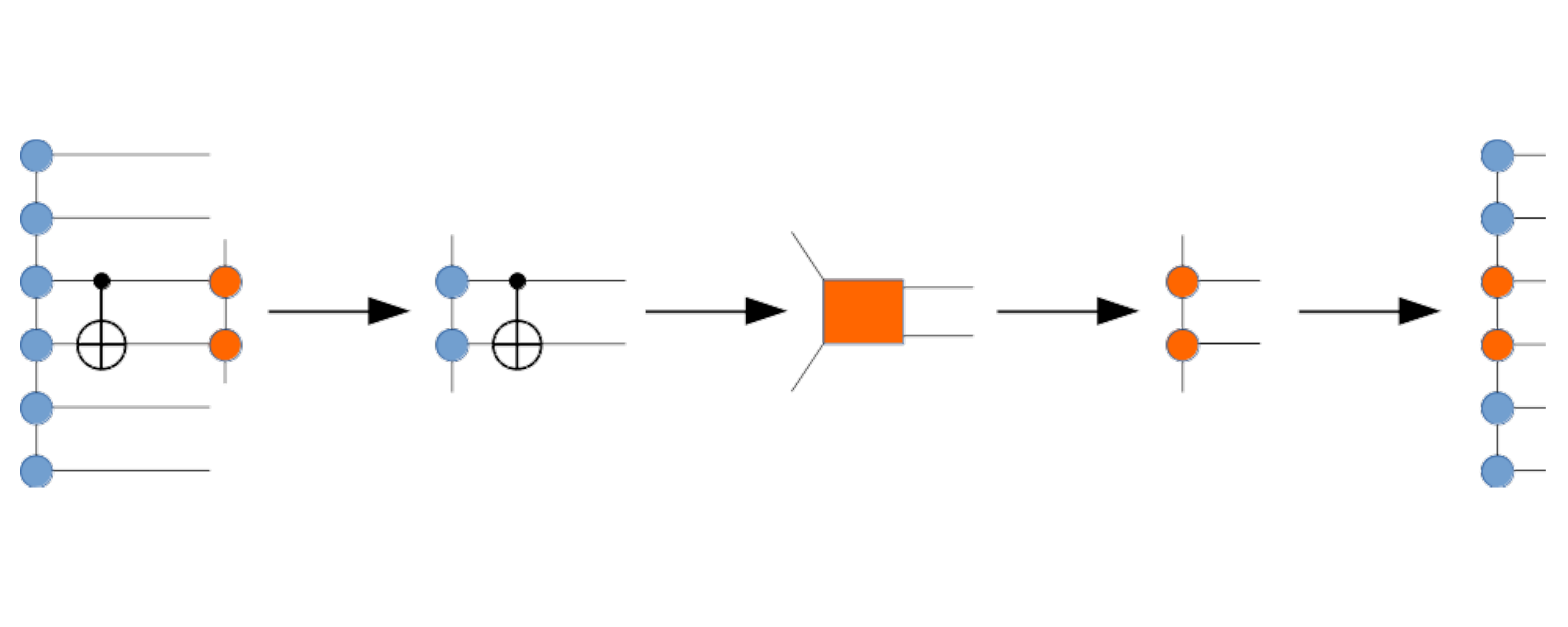}
\caption{Graphical illustration of an accelerated evaluation of the action of a two-body gate on a pair of adjacent qubits in the matrix-product state representation.}
\label{fig:mpssvd}
\end{figure}

In cases where an accelerated gate application is possible (for example, a 2-body gate is applied to the adjacent qubits in the MPS-factorized wave-function), one can restrict the update procedure only to the tensor factors directly affected by the gate action. In case of MPS factorization, in order to apply a 2-body gate to two adjacent qubits one can contract the gate tensor with the two MPS tensors representing the affected qubits and then perform the singular value decomposition (SVD) on the tensor-result, thus obtaining the new (updated) MPS tensors as illustrated in Figure \ref{fig:mpssvd}.

The above general algorithm demonstrates the procedure used by TNQVM for approximate simulation of quantum circuits based on the tensor network factorization. For the sake of completeness, we should also mention quantum circuit simulators which use tensor representations for a brute-force simulation of quantum circuits with no approximations \cite{Pednault2017, fried2017qtorch}. This is different from our approach which is based on the explicit factorization of the multi-qubit wave-function tensor. In these other tensor-based schemes the entire quantum circuit as a collection of gate tensors is considered as a tensor network which is subsequently contracted over in order to compute observables or evaluate output probability distributions. In Ref.~\citenum{Pednault2017}, a clever tensor slicing technique was introduced that avoided the evaluation of the full wave-function tensor, thus reducing the memory footprint and bypassing the existing 45-qubit limit on large-scale HPC systems. Yet, despite enabling simulations of somewhat larger qubit counts, this technique does not lift the asymptotic bounds of the exact simulation cost.
\par
\section{Quantum Virtual Machines}
\label{sec:qvm}
In order to evaluate the correctness of a quantum program and its implementation via a decomposition into primitive gate operations, it is necessary to model both the conventional computing and quantum computing elements of the system architecture. In particular, it is necessary to expose the interface to the available instruction set architecture (ISA) and methods to support quantum program execution, scheduling, and layout. There are currently many different technologies available for testing and evaluating quantum processing units, and each of these technologies presents different ISAs and methods for program execution \cite{Britt2017ISA}.
\par
\begin{figure}[ht]
\centering
\includegraphics[width=2.0in]{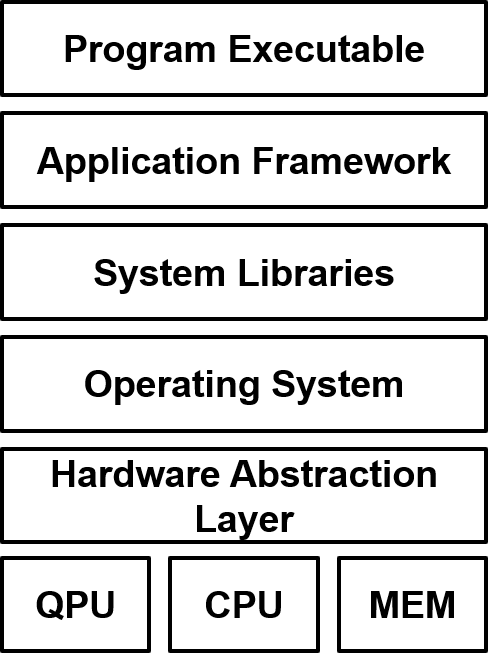}
\caption{A schematic design how a quantum virtual machine (QVM) manages access to an underlying QPU through the hardware abstraction layer. A program binary exists within an application framework that accesses system resources through libraries. Library calls are managed by the host operating system, which manages and schedules requests to access hardware devices including attached QPUs. The hardware abstraction layer (HAL) provides a portable interface by which these requests are made to the underlying QPU technology.}
\label{fig:qvm}
\end{figure}

As shown in Fig.~\ref{fig:qvm}, a quantum virtual machine (QVM) provides a portable abstraction of technology-specific details for a broad variety of heterogeneous quantum-classical computing architectures. The hardware abstraction layer (HAL) defines a portable interface by which the underlying quantum processor technology as well as other hardware components such as memory are exposed to system libraries, runtimes and drivers running on the host conventional computer. The implementation of the HAL provides an explicit translation of quantum program instructions into native, hardware-specific syntax, which may be subsequently executed by the underlying quantum processor. The HAL serves as a convenience to ensure portability of programs across different QPU platforms, while the QVM encapsulates the environment in which applications can be developed independently from explicit knowledge of QPU details. This environment is provided by the integration of the HAL with programming tools, libraries, and frameworks as well as the host operating system.
\par
Application performance within a QVM depends strongly on the efficiency with which host programs are translated into hardware-specific instructions. This includes the communication overhead between the HAL and hardware layers as well as the overhead costs for managing these interactions by the host operating system. Both algorithmic and hardware designs impact this performance by deciding when and how to allocate computational burden to specific devices. Presently, there is an emphasis on the development and validation of hybrid programs, which loosely integrates quantum processing with conventional post-processing tasks. This algorithmic design introduces a requirement for transferring memory buffers between the host and QPU systems. Memory management therefore becomes an important task for application behavior. While current QPUs are often accessed remotely through network interfaces, long-term improvements in application performance will require fine grain control over memory management.  
\section{Tensor Network Quantum Virtual Machine}
Our implementation of a QVM presented in this work is based on a previously developed hybrid quantum-classical programming framework, called XACC \cite{xaccarxiv}, combined with a quantum circuit simulator that uses tensor network theory for compressing the multi-qubit wave-function. We provide an overview of the Tensor Network Quantum Virtual Machine (TNQVM) and its applications, including its software architecture and integration with the XACC programming framework. Since XACC integrates directly with TNQVM, compiled programs can in principle be verified instantaneously on any classical computer including workstations as well as HPC clusters and supercomputers. The support of different classical computer architectures (single-core, multi-core, GPU, distributed) for performing numerical simulations is achieved by interchangeability of the numerical backends in our TNQVM simulator. These backends are numerical tensor algebra libraries which perform all underlying tensor computations on a supported classical computer. In this work, we detail the HAL implementation of TNQVM using ITensor \cite{itensor} for serial simulations, with some example applications demonstrating the utility of TNQVM. We also sketch some details on the upcoming ExaTENSOR backend that will enable large-scale quantum circuit simulations on homo- and heterogeneous HPC systems. Independent verification of hybrid programs within TNQVM provides an increased confidence in the use of these codes to characterize and validate actual QPUs.
\subsection{XACC}
The eXtreme-scale ACCelerator programming model (XACC) has been specifically designed for enabling near-term quantum acceleration within existing classical high-performance computing applications and workflows \cite{xaccarxiv, mccaskeyicrc}. This programming model and associated open-source reference implementation follow the traditional co-processor model, akin to OpenCL or CUDA for GPUs, but takes into account the subtleties and complexities arising from the interplay between classical and quantum hardware. XACC provides a high-level application programming interface (API) that enables classical applications to offload quantum programs (represented as quantum kernels, similar in structure to GPU kernels) to an attached quantum accelerator in a manner that is agnostic to both the quantum programming language and the quantum hardware. Hardware agnosticism enables quantum code portability and also aids in benchmarking, verification and validation, and performance studies for a wide array of virtual (simulators) and physical quantum platforms. 

To achieve language and hardware interoperability, XACC defines three important abstractions: the quantum intermediate representation (IR), compilers, and accelerators. XACC compiler implementations map quantum source code to the IR -- the in-memory object key to integrating of a diverse set of languages to a diverse set of hardware. IR instances (and therefore compiled kernels) are executed by realizations of the accelerator concept, which defines an interface for injecting physical or virtual quantum hardware. Accelerators take this IR as input and delegate execution to vendor-supplied APIs for the QPU, or an associated API for a simulator. This forms the hardware abstraction layer, or abstract device driver, necessary for a general quantum (virtual) machine.

\begin{figure}[htb!]
\centering
\includegraphics[width=.45\textwidth]{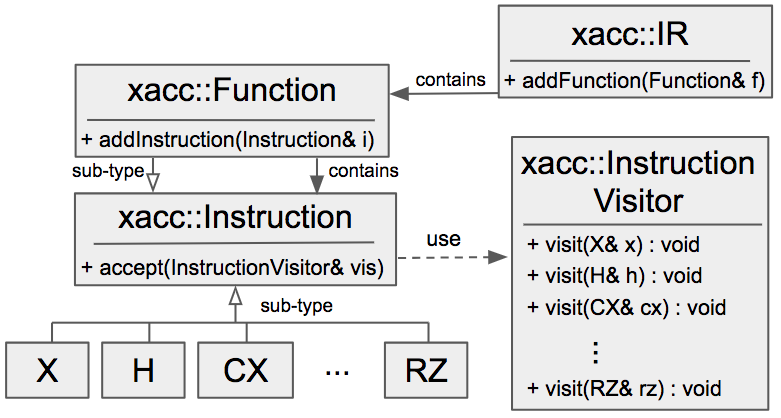}
\caption{Architecture of the XACC intermediate representation demonstrating sub-type extensibility for instructions, and the associated instruction visitor abstraction, enabling runtime-extension of concrete instruction functionality.}
\label{fig:xacc-visitor}
\end{figure}

The IR itself can be further decomposed into instruction and function concepts, with instructions forming the foundation of the IR infrastructure and functions serving as compositions of instructions (see Figure \ref{fig:xacc-visitor}). Each instruction exposes a unique name and the set of qubits that it operates on. Functions are a sub-type of the instruction abstraction that can contain further instructions. This setup, the familiar composite design pattern \cite{composite}, forms an {\it n-ary} tree of instructions where function instances serve as nodes and concrete instructions instances serve as leaves. 

Operating on this tree and executing program instructions is a simple pre-order traversal on the IR tree. In order to enhance this tree of instructions with additional functionality, XACC provides a dynamic double-dispatch mechanism, specifically an implementation of the familiar visitor pattern \cite{gof}. The visitor pattern provides a mechanism for adding virtual functions to a hierarchy of common data structures dynamically, at runtime, and without modifying the underlying type. This is accomplished via the introduction of a visitor type that exposes a public set of visit functions, each one taking a single argument that is a concrete sub-type of the hierarchical data structure composition (see Figure \ref{fig:xacc-visitor}). For gate model quantum computing, XACC exposes a visitor class that exposes a visit method for all concrete gate instructions (X, H, RZ, CX, etc...). All instructions expose an \texttt{accept} method that takes as input a general visitor instance, and invokes the appropriate visit method on the visitor through double-dispatch. XACC instruction visitors thereby provide an extensible mechanism for dynamically operating on, analyzing, and transforming compiled IR instances at runtime. 

\subsection{Tensor Network Accelerator and Instruction Visitors}
The integration of a tensor network quantum circuit simulator with XACC can be accomplished through extensions of appropriate XACC concepts. In essence, this is an extension of the quantum virtual machine hardware abstraction layer that enables existing high-level programs and libraries to target a new virtual hardware instance. Injecting new simulators into the XACC framework requires a new implementation of the accelerator concept. Enabling that simulator to be extensible in the type of tensor networks, algorithmic execution, and the library backend requires different mappings of the IR to appropriate simulation data structures. This can be accomplished through individual implementations of the instruction visitor concept. 

Our open-source implementation of the Tensor Network Quantum Virtual Machine (TNQVM) library extends the XACC accelerator concept with a new derived class that simulates pure-state quantum computation via tensor network theory \cite{tnqvm-github}. This library provides the TNAccelerator (Tensor Network Accelerator) that exposes an \texttt{execute} method that takes as input the XACC IR function to be executed. Generality in the tensor network graph structure and the simulation algorithm is enabled through appropriate implementations of the instruction visitor concept. For example, an instruction visitor can be implemented to map the incoming XACC IR tree to tensor operations on a matrix product state (MPS) ansatz. Walking the IR tree via pre-order traversal and invoking the instruction visitor \texttt{accept} mechanism on each instruction triggers invocation of the appropriate visit function via double dispatch. The implementation of these visit methods provides an extensible mechanism for performing instruction-specific tensor operations on a specific tensor network graph structure. 

Furthermore, this visitor extension mechanism can be leveraged to not only provide new tensor network structures and operations, but also provide the means to leverage different tensor algebra backend libraries, and therefore introduce a classical parallel execution context. Different visitor implementations may provide a strictly serial simulation approach, while others can enable a massively parallel or heterogeneous simulation approach (incorporating the Message Passing Interface, OpenMP, and/or GPU acceleration via CUDA or OpenCL).

To date we have implemented two instruction visitor backends for the TNQVM and the TNAccelerator. We have leveraged the ITensor library \cite{itensor} to provide a serial matrix product state simulator, and the ExaTENSOR library from the Oak Ridge Leadership Computing Facility (OLCF) to provide a matrix product state simulator that leverages MPI, OpenMP and CUDA for distributed parallel execution on GPU-accelerated heterogeneous HPC platforms. However, the ExaTENSOR library is currently undergoing final testing before its public release, thus it has not been utilized yet as a fully functional backend of TNQVM. Nevertheless, we will provide some details on the ExaTENSOR backend below.

\subsubsection{ITensor MPS Implementation}
The ITensor MPS instruction visitor implementation provides a mechanism for the simulation of an $N$-qubit wavefunction via a matrix product state tensor network decomposition. The MPS provides a way to restrict the entanglement entropy through SVD and associated truncation of Schmidt coefficients to reduce the overall Schmidt rank. With these MPS states, we need $O(n\chi^2)$ numbers to represent $n$ qubits, where $\chi$ is the largest Schmidt rank we keep. As long as $\chi$ is not too large (grows polynomially with system size), the space complexity is feasible for classical simulation. For example, if the quantum register is used to store the gapped ground states of systems with local interactions, we can simulate larger number of qubits and still adequately approximate the wavefunction by keeping $\chi$ small enough.

Our ITensor MPS visitor implementation begins by initializing a matrix product state tensor network using the serial tensor data structures provided by the ITensor library\cite{itensor}. Simulation of the compiled IR program is run through a pre-order tree traversal of the instruction tree. At each leaf of this tree (a concrete instruction), the \texttt{accept} method on the instruction is invoked (see Figure \ref{fig:xacc-visitor}) which dispatches a call to the correct \texttt{visit} method of the instruction visitor.

At this point, the appropriate gate tensor is contracted into the MPS representation, which maps onto itself under local quantum gates. Updating the MPS according to two-body entanglers involves two-qubit gates which act on two rank-3 tensors, and the full contraction results in a rank-4 tensor. We maintain the MPS structure by decomposing the rank-4 tensor into two rank-3 tensors and a diagonal matrix between them. Note that when the two qubits are not adjacent we apply SWAP gates on intermediary qubits to bring them together. The gate is then applied and reverse SWAPs bring the qubits back to the original positions. Otherwise, applying a gate to non-adjacent qubits would modify the underlying graph topology, complicating future evolution by adding an non-local loop in the tensor network.

The SVD is used to return the resulting rank-4 tensor to the canonical MPS form ($n$ rank-3 tensors and $n-1$ diagonal matrices), with the singular values below a cutoff threshold $\epsilon$ (e.g., default is $\epsilon=10^{-4}$) being truncated. The truncation over subspaces supporting exponentially small components of the wave-function allows our MPS-based TNQVM simulate large numbers of qubits, contingent on some slowly growing entanglement properties. Examples and discussion may be found in the demonstrations in Sec.~\ref{sec:demonstration}.




\subsubsection{ExaTENSOR MPS Implementation}
The ExaTENSOR numerical tensor algebra backend will enable larger-scale TNQVM quantum circuit simulations on GPU-enabled and other accelerated as well as conventional multicore HPC platforms. ExaTENSOR stores tensors in distributed memory (on multiple/many nodes) as a generally sparse collection of tensor slices in a hierarchical fashion. Such distributed tensor storage lifts the memory limitations pertinent to a single node, thus extending the maximal number of simulated qubits. Although we currently target the (distributed) MPS implementation, ExaTENSOR also provides a generic tensor network builder that can be used for constructing an arbitrary tensor network. The ExaTENSOR MPS visitor implementation provides a constructor that creates the MPS representation of the simulated multi-qubit wave-function (all constituent MPS tensors are distributed now). Then the XACC IR tree traversal invokes ExaTENSOR MPS \texttt{visit} method for each traversed node (instruction). The \texttt{visit} method implements lazy visiting, namely it only caches the corresponding instruction (gate) in the instruction cache of the ExaTENSOR MPS visitor. At some point, once the instruction cache has enough work to perform, the \texttt{evaluate} method of the ExaTENSOR visitor is invoked which implements the generic gate action algorithm shown in Section II. Specifically, it allocates the output MPS tensor network, that is, the result of the action of the cached gates on the input MPS tensor network. Then it creates the inner product (closed) tensor network by joining the gate tensors to the input MPS tensor network, subsequently closing it with the output tensor network (see Figure \ref{fig:tnalg}). This closed tensor network is a scalar whose value needs to be maximized. The ExaTENSOR MPS visitor will utilize the standard gradient descent algorithm by evaluating the gradient with respect to each tensor constituting the output tensor network. Each of these gradients is an open tensor network itself that needs to be fully contracted. Importantly, the computational cost of this contraction of many tensors strongly depends on the order in which the pairwise tensor contractions are performed. Finding the optimal tensor contraction sequence is an NP-hard problem. Instead, ExaTENSOR implements a heuristic algorithm that delivers the best found sequence of pairwise tensor contractions in a reasonable amount of time (subseconds). Then this pseudo-optimal sequence of pairwise tensor contractions is cached for a subsequent reuse, if needed. Given the sequence of pairwise tensor contractions, the ExaTENSOR library will numerically evaluate all of them and return the gradients that will subsequently be used for updating the output tensor network tensors, until the optimized inner product scalar reaches the desired value. In case it does not reach the desired value, the tensors constituting the output tensor network are reallocated with increased dimensions of the auxiliary spaces and the entire procedure is repeated. At this point, the early prototype implementation of the ExaTENSOR MPS visitor in TNQVM is based on the single-node version of the ExaTENSOR library and we are currently finishing the integration of the TNQVM with the distributed version of the ExaTENSOR library as well as performing the final testing of the ExaTENSOR library itself before the public release scheduled later this year.

\section{Demonstration}
\label{sec:demonstration}
Here we demonstrate the utility of TNQVM by describing the overall memory scaling of our matrix product state TNQCM for varying levels of entanglement and system size. Our demonstrations show how TNQVM can be leveraged to validating hybrid quantum-classical programming models. Specifically we focus on random circuit simulations and the variational quantum eigensolver (VQE) hybrid algorithm. 

\subsection{Profiling Random Circuit Simulations with MPS}
We demonstrate the improved resource cost of representing quantum states ($O(n\chi^2)$ vs $O(2^n)$) with TNQVM by using an MPS formulation and by profiling the memory usage of simulating randomly generated circuits. We vary the entanglement structure of our random circuits by constructing time slices defined as \emph{rounds}. The first round begins with a layer of Hadamard operations on all qubits, followed by a layer of single qubit gates (Pauli gates and other general rotations), followed by a set of nearest-neighbor CNOT entangling operations. Multiple rounds constitute multiple iterations of generating these layers (excluding the Hadamards, which only appear in the first layer). Clearly, later rounds add layers of entangling CNOT operations and therefore generate states with a more complicated entanglement structure. 
\begin{figure}[ht]
\centering
\includegraphics[width=\columnwidth]{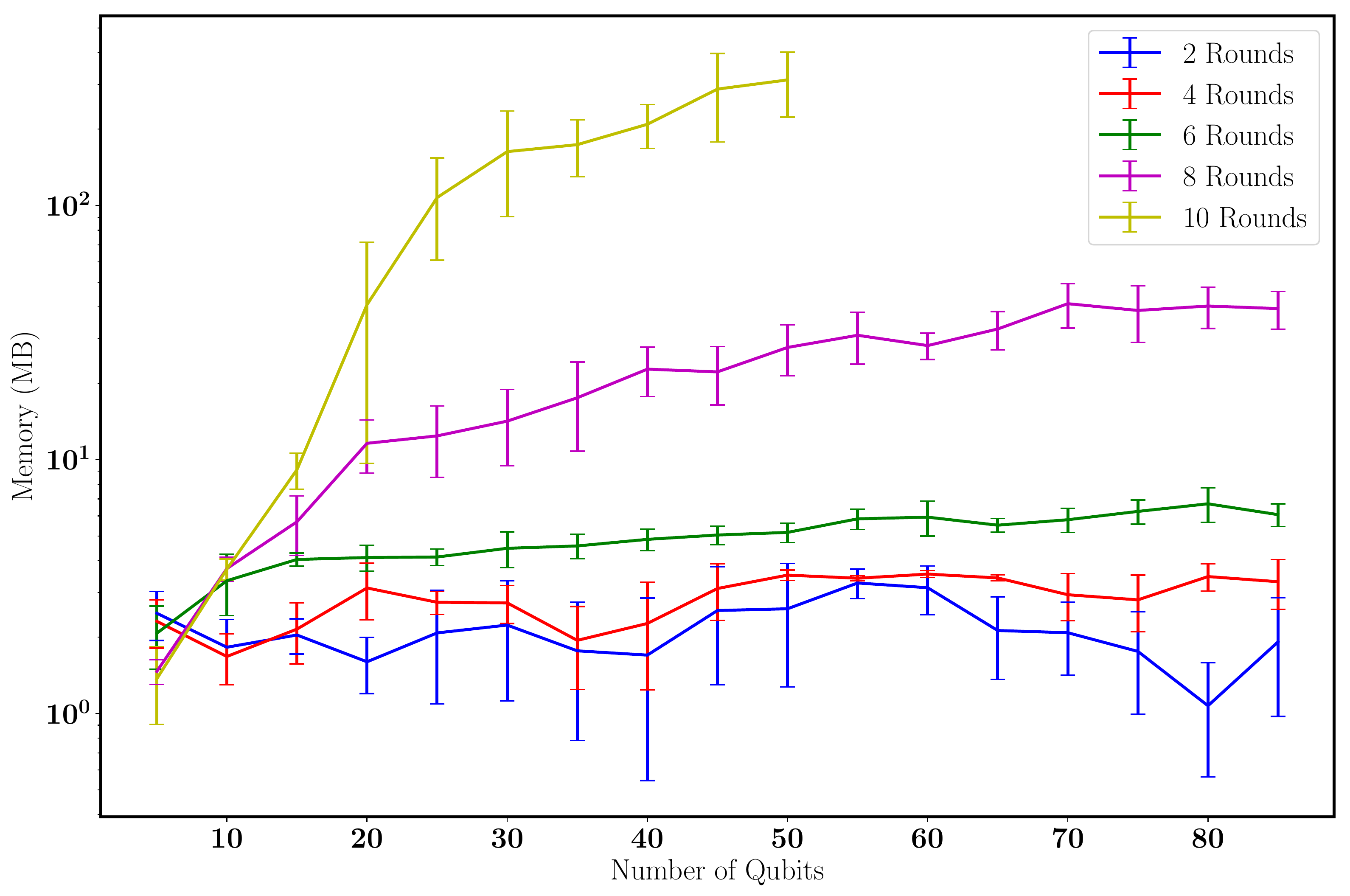}
\caption{Memory usage as a function of the number of rounds (circuit depth) and with increasing number of qubits. Memory usage is constant for a small number of rounds but rapidly increases as the total circuit depth and number of qubits increases. }
\label{fig:profiling}
\end{figure}

We generate these random circuits for $5$ through $85$ qubits in increments of $5$, and for numbers of rounds ranging from $2$ through $10$ in increments of $2$. For each $(round, n-qubits)$ pair, we generate 10 random circuits, compute the heap memory usage, and compute the mean and standard deviation of the memory usage. The results are plotted in Figure \ref{fig:profiling}. For lightly-entangled systems (i.e. those generated by a small number of random rounds) we see that the MPS structure is able to encode the wavefunction of the system efficiently with a small cost. For example, for only two rounds the maximum bond dimension is $\chi=4$, which is independent of system size.   As we increase the entanglement in our random circuits, the computational cost of the MPS simulations increases exponentially. This is because circuits we have sampled from are designed to saturate exponentially increase the entanglement which saturates at entanglement described by $\chi_{max} = 2^{n/2}$ of an $n$-qubit system undergoing $m>n$ random rounds\cite{Boixo2018}.

\subsection{Variational Quantum Eigensolver}
\begin{figure}[!htb]
 \begin{minipage}{0.45\textwidth}
  \begin{listing}[H]
\begin{Verbatim}[commandchars=\\\{\}]
\PYGdefault{n}{h2\PYGdefaultZus{}src} \PYGdefault{o}{=} \PYGdefault{l+s+s2}{\PYGdefaultZdq{}\PYGdefaultZdq{}\PYGdefaultZdq{}}
\PYGdefault{l+s+s2}{\PYGdefaultZus{}\PYGdefaultZus{}qpu\PYGdefaultZus{}\PYGdefaultZus{} ansatz(AcceleratorBuffer b,}
\PYGdefault{l+s+s2}{                            double t0) \PYGdefaultZob{}}
\PYGdefault{l+s+s2}{  RX(3.1415926) 0}
\PYGdefault{l+s+s2}{  RY(1.57079) 1}
\PYGdefault{l+s+s2}{  RX(7.85397) 0}
\PYGdefault{l+s+s2}{  CNOT 1 0}
\PYGdefault{l+s+s2}{  RZ(t0) 0}
\PYGdefault{l+s+s2}{  CNOT 1 0}
\PYGdefault{l+s+s2}{  RY(7.8539752) 1}
\PYGdefault{l+s+s2}{  RX(1.57079) 0}
\PYGdefault{l+s+s2}{\PYGdefaultZcb{}}
\PYGdefault{l+s+s2}{\PYGdefaultZus{}\PYGdefaultZus{}qpu\PYGdefaultZus{}\PYGdefaultZus{} term0(AcceleratorBuffer b, double t0) \PYGdefaultZob{}}
\PYGdefault{l+s+s2}{  ansatz(b, t0)}
\PYGdefault{l+s+s2}{  MEASURE 0 [0]}
\PYGdefault{l+s+s2}{\PYGdefaultZcb{}}
\PYGdefault{l+s+s2}{... (rest of measurement kernels)}
\PYGdefault{l+s+s2}{\PYGdefaultZdq{}\PYGdefaultZdq{}\PYGdefaultZdq{}}
\PYGdefault{n}{qpu} \PYGdefault{o}{=} \PYGdefault{n}{xacc}\PYGdefault{o}{.}\PYGdefault{n}{getAccelerator}\PYGdefault{p}{(}\PYGdefault{l+s+s1}{\PYGdefaultZsq{}tnqvm\PYGdefaultZsq{}}\PYGdefault{p}{)}
\PYGdefault{n+nb}{buffer} \PYGdefault{o}{=} \PYGdefault{n}{qpu}\PYGdefault{o}{.}\PYGdefault{n}{createBuffer}\PYGdefault{p}{(}\PYGdefault{l+s+s1}{\PYGdefaultZsq{}q\PYGdefaultZsq{}}\PYGdefault{p}{,}\PYGdefault{l+m+mi}{2}\PYGdefault{p}{)}

\PYGdefault{n}{p} \PYGdefault{o}{=} \PYGdefault{n}{xacc}\PYGdefault{o}{.}\PYGdefault{n}{Program}\PYGdefault{p}{(}\PYGdefault{n}{qpu}\PYGdefault{p}{,} \PYGdefault{n}{h2\PYGdefaultZus{}src}\PYGdefault{p}{)}
\PYGdefault{n}{p}\PYGdefault{o}{.}\PYGdefault{n}{build}\PYGdefault{p}{()}
\PYGdefault{n}{kernels} \PYGdefault{o}{=} \PYGdefault{n}{p}\PYGdefault{o}{.}\PYGdefault{n}{getKernels}\PYGdefault{p}{()}

\PYGdefault{k}{for} \PYGdefault{n}{t0} \PYGdefault{o+ow}{in} \PYGdefault{n}{np}\PYGdefault{o}{.}\PYGdefault{n}{linspace}\PYGdefault{p}{(}\PYGdefault{o}{\PYGdefaultZhy{}}\PYGdefault{n}{np}\PYGdefault{o}{.}\PYGdefault{n}{pi}\PYGdefault{p}{,}\PYGdefault{n}{np}\PYGdefault{o}{.}\PYGdefault{n}{pi}\PYGdefault{p}{,}\PYGdefault{l+m+mi}{100}\PYGdefault{p}{):}
    \PYGdefault{k}{for} \PYGdefault{n}{k} \PYGdefault{o+ow}{in} \PYGdefault{n}{kernels}\PYGdefault{p}{[}\PYGdefault{l+m+mi}{1}\PYGdefault{p}{:]:}
        \PYGdefault{n}{k}\PYGdefault{o}{.}\PYGdefault{n}{execute}\PYGdefault{p}{(}\PYGdefault{n+nb}{buffer}\PYGdefault{p}{,}
                \PYGdefault{p}{[}\PYGdefault{n}{xacc}\PYGdefault{o}{.}\PYGdefault{n}{InstructionParameter}\PYGdefault{p}{(}\PYGdefault{n}{t0}\PYGdefault{p}{)])}
\end{Verbatim}

\end{listing}
\end{minipage}
 \caption{XACC program compiling and executing the variational quantum eigensolver for the $H_2$ molecule.}
\end{figure}
Finally, we demonstrate the utility of our tensor network simulation XACC Accelerator backend (the TNQVM library) in validating quantum-classical algorithms. It is this rapid feedback mechanism that is critical to understanding intended algorithmic results, and enables confidence in the programming of larger systems. Here we demonstrate this programmability and its verification and validation through a simple simulation of diatomic hydrogen via the variational quantum eigensolver algorithm. The quantum-classical program is shown in the listing below leveraging the TNQVM library.

This code listing demonstrates the integration of XACC and our tensor network accelerator implementation. The code shows how to program, compile, and execute the VQE algorithm to compute expectation values for the simplified (symmetry-reduced), two qubit $H_2$ Hamiltonian (see \cite{babbush_scalable_sim}). We start off by defining the quantum source code as XACC quantum kernels (note - we have left out a few measurement kernels for brevity). Each of these kernels is parameterized by a single \texttt{double} representing the variational parameter for the problem ansatz circuit (the \texttt{ansatz} kernel in the \texttt{h2\_src} string). Integration with the TNQVM simulation library is done through a public XACC API function (\texttt{getAccelerator}). This accelerator reference is used to compile the program and get reference to executable kernels that delegate work to the TN Accelerator. We then loop over all $\theta$ and compute the expectation values for each Hamiltonian measurement term. Notice that this execution mechanism is agnostic to the accelerator sub-type. This provides a way to quickly swap between validation and verification with TNQVM, and physical hardware execution on quantum computers from IBM, Rigetti, etc.

\section{Conclusion}
In this work we have discussed the concept of a general quantum virtual machine and introduced a concrete implementation of the QVM that enables quantum-classical programming with validation through an extensible tensor network quantum circuit simulator (TNQVM). We have discussed the applicability and scalability of a matrix product state backend implementation for TNQVM and discussed the role of TNQVM in benchmarking quantum algorithms and hybrid quantum-classical applications including random circuit sequences used in quantum supremacy\cite{Boixo2018} and the variational quantum eigensolver\cite{Peruzzo2014}. We have chosen a tensor network based quantum virtual machine due to the complexity reduction such a formalism provides for a broad range of problems. In general TNQVM enables large-scale simulation of quantum circuits which generate states characterized by short-range entanglement. Studying systems with long-range entanglement interactions will require further developments in implementing more advanced tensor network decomposition types. We plan to investigate the applicability of the tree tensor network and the multiscale entanglement renormalization ansatz in future work, in an effort to scale simulation capabilities to a larger number of qubits.

\section*{Acknowledgements}
This work has been supported by the Laboratory Directed Research and Development Program of Oak Ridge National Laboratory and the US Department of Energy (DOE) Early Career Research Program. This research used resources of the Oak Ridge Leadership Computing Facility, which is a DOE Office of Science User Facility supported under Contract DE-AC05-00OR22725. This manuscript has been authored by UT-Battelle, LLC, under contract DEAC0500OR22725 with DOE. The US government retains and the publisher, by accepting the article for publication, acknowledges that the US government retains a nonexclusive, paid-up, irrevocable, worldwide license to publish or reproduce the published form of this manuscript, or allow others to do so, for US government purposes. DOE will provide public access to these results of federally sponsored research in accordance with the DOE Public Access Plan.

\bibliographystyle{apsrev}
\bibliography{main}

\end{document}